\documentstyle[12pt]{article}
\begin{document}
\textwidth = 14 truecm
\textheight = 20 truecm
\baselineskip25pt
\begin{center}
{\large \bf Bounds on R-parity violating SUSY Yukawa couplings from 
semileptonic decays of baryons} 
\end{center}
\vspace{1cm}
\begin{center}
Farida Tahir \footnote{E-mail: anwar@sgs.sdnpk.undp.org}
, M. Sadiq, M. Anwar Mughal, 
K. Ahmed \footnote{E-mail: kahmed@hep-qau.sdnpk.undp.org} \\ 
 {\em Department of Physics, Quaid-i-Azam University, Islamabad, 
Pakistan}\\
\end{center}
\newcommand{\be}{\begin{equation}}
\newcommand{\ee}{\end{equation}}
\newcommand{\bea}{\begin{eqnarray}}
\newcommand{\eea}{\end{eqnarray}}
\newcommand{\btab}{\begin{tabular}}
\newcommand{\etab}{\end{tabular}}
\newcommand{\bef}{\begin{figure}}
\newcommand{\eef}{\end{figure}}
\newcommand{\bt}{\begin{table}}
\newcommand{\et}{\end{table}}
\newcommand{\ben}{\begin{enumerate}}
\newcommand{\een}{\end{enumerate}}
\newcommand{\ba}{\begin{array}}
\newcommand{\ea}{\end{array}}
\newcommand{\bei}{\begin{itemize}}
\newcommand{\eei}{\end{itemize}}
\newcommand{\al}{\alpha}
\newcommand{\ct}{\cite}
\newcommand{\lb}{\label}
\title{}
\vspace{1.5in}
\begin{abstract}
We consider tree-level corrections to hypercharge changing 
semileptonic 
decays of certain baryons induced by a minimal supersymmetric 
standard 
model 
with explicitly broken $R$-parity via $L$-violation. This study leads 
to a 
new set of constraints on the products of couplings arising from the 
LQd$^c$ 
operator of the superpotential. 
\end{abstract}
\newpage
Lepton number ($L$) is conserved in the Standard Model (SM) of 
electroweak
interactions; discovery of $L$-violation would be recognized as a
manifestation of physics beyond the SM. Among the possible extensions 
of the
SM, supersymmetry (SUSY) in its minimal form known as the minimal
supersymmetric SM (MSSM) could be the most attractive possibility of 
$L$%
-violation. In general, the MSSM could contain renormalizable $L$-
violating
interactions as well as the baryon number ($B$) violating interaction 
which
are described by the superpotential \ct{1}

\begin{equation}
f^{\Delta L\neq 0}=\frac 12\lambda _{ijk}\left[ \mbox{L}_i \mbox{L}_j 
\right] 
\mbox{e}_k^c+\lambda _{ijk}^{^{\prime }} \mbox{L}_i \mbox{Q}_j 
\mbox{d}%
_k^c+\lambda _{ijk}^{^{\prime \prime }} \mbox{u}_i^c \mbox{d}_j^c 
\mbox{d}_k^c,
\lb{eq:1}
\end{equation}
where $i,j,k$ are generation indices, L, Q are the lepton and quark
left-handed doublets and $e^c,$ $u^c\left( d^c\right) $ are the charge
conjugates of the right-handed lepton and {\it up(down) }quark 
singlets,
respectively. $\lambda _{ijk},\lambda _{ijk}^{^{\prime }}$ and 
$\lambda
_{ijk}^{^{\prime \prime }}$ are the Yukawa couplings. $\lambda 
_{ijk}$ 
and $%
\lambda _{ijk}^{^{\prime \prime }}$ are antisymmetric in $i,j$ and 
$j,k$
indices respectively. Renormalizable $L$- and $B$- violating gauge 
invariant
interactions are the characteristic features of the SUSY theory. The
doubling of particle spectrum so arising leads to the introduction of
squarks and sleptons which couple to standard fermions through $L$- 
and 
$B$-
violating couplings. There has been a lot of work in the recent years 
on the
effects associated with the $R$-breaking couplings in (\ref{eq:1}). 
$R$-
breaking
interactions contribute to various low-energy processes through the 
virtual
exchange of sparticles which in turn put bounds on the $R$-breaking
couplings. Among many papers constraining individual couplings 
include 
the
constraints from atomic parity violation and $eD$-asymmetry, $\nu 
_\mu 
$ DIS
\ct{2}, $\left( \nu _0\right) _{\beta \beta }$ decay \ct{3}, 
neutrino mass \ct{4},
decays of $K^{+},$ top quark \ct{5} and $Z$ \ct{6}. More recent 
reviews 
on the
effects of $R$-breaking with updated bounds on the couplings are 
given 
in
Ref. \ct{7}. Phenomenologies of these couplings at various colliders 
have also
been investigated recently \ct{8}. Note that together, the couplings 
in 
the $L$%
- and $B$ -violating terms in (\ref{eq:1}), 
lead to a rapid proton decay \ct{9}. The
standard remedy for this is enforcing a new symmetry called the $R$-
parity
\ct{1,9} which is defined as $R=\left( -\right) ^{3B+L+2S}$, with a 
value $+1$
for ordinary particles and $-1$ for their SUSY partners. As an 
alternate a
phenomenology of $R$ -breaking due to explicit $L$-violation is 
equally
viable that can also provide a natural protection to proton by 
disallowing $%
B$-violating interaction.

While establishing the $R$-violating operators in 
(\ref{eq:1}), it is generally
assumed that one $R$-violating operator is dominant at a time. Under 
this
assumption sometime stronger bounds on the products of $R$-violating
operators are neglected. There are already some papers where bounds 
on 
the
products $\lambda ^{^{\prime }}\lambda ^{^{\prime }}$ were derived 
from
various processes. Some important bounds include a bound on the 
product 
$%
\lambda _{113}^{^{\prime }}\lambda _{131}^{^{\prime }}$ from the
neutrinoless double beta decay [2], $\lambda _{i13}^{^{\prime 
}}\lambda
_{i31}^{^{\prime }}$ and $\lambda _{i12}^{^{\prime }}\lambda
_{121}^{^{\prime }}$ from neutral mesons mixing, $\lambda 
_{1j1}^{^{\prime
}}\lambda _{2j1}^{^{\prime }}$, $\lambda _{11k}^{^{\prime }}\lambda
_{21k}^{^{\prime }}$ from muonium conversion, and $\lambda 
_{1j1}^{^{\prime
}}\lambda _{2j2}^{^{\prime }}$ 
from the flavour changing decays of $K$ \ct{11}.
The decays of $\tau $ into lepton and vector/pseudoscalar mesons puts 
constraints
on $\lambda _{31k}^{^{\prime }}\lambda _{11k}^{^{\prime }}$ and 
$\lambda
_{31k}^{^{\prime }}\lambda _{21k}^{^{\prime }}$ \ct{12}. Recently new 
limits on
various $\lambda ^{^{\prime }}\lambda ^{^{\prime }}$ products have 
been
derived from the lepton-flavour violating 
$Z$-decays \ct{13} and from the
consideration of $\Delta S=2$ 
and $\Delta B=2$ box graphs \ct{14}.

In this note we consider the hypercharge changing semileptonic decays 
of
baryons to derive a new set of bounds on certain products of $\lambda
^{^{\prime }}$ couplings by calculating the $L-$ violating 
contribution 
to
such decays. The relevant Lagrangian in terms of component fields is 
given by

\begin{equation}
{\em L}_{\lambda ^{^{\prime }}}=\lambda _{ijk}^{^{\prime }}\left[ 
\begin{array}{c}
\overline{d}_{kR}\nu _{iL}\widetilde{d}_{jL}+\mbox{ 
}\overline{d}_{kR}d_{iL}%
\widetilde{\nu }_{jL}+\mbox{ }\overline{\left( \nu _{iL}\right) 
^c}d_{jL}%
\widetilde{d}_{kR}^{*} \\ 
-\overline{d}_{kR}e_{iL}\widetilde{u}_{jL}-\mbox{ 
}\overline{d}_{kR}u_{jL}%
\widetilde{e}_{iL}-\overline{\left( e_{iL}\right) 
^c}u_{jL}\widetilde{d}%
_{kR}^{*}
\end{array}
\right] +h.c. 
\lb{eq:2} 
\end{equation}
Evidently, the above $L$-violating couplings manifest themselves only 
when
the two non-spectator quarks form an $SU\left( 2\right) _L$ doublets. 
To be
explicit we consider a typical hypercharge changing semileptonic 
baryonic
decay: $B^{}\longrightarrow B^{^{\prime }}l\overline{\nu }_l$ $\left(
l=e,\mu \right) $ with a baryon in the final state. We have chosen 
$\Delta
Y=1$ decay modes as they involve the products of LQd$^c$ operators. 
All 
such
decays can be represented by quark subprocess $s\longrightarrow 
ul\overline{%
\nu }$. These decays proceed in the SM by a tree-level $W$-exchange 
Feynman
graph. In the presence of interaction (\ref{eq:2}) an additional 
squark-
exchange
diagram contributes to these decays. The current structure of this new
interaction is given by
\begin{equation}
{\em L}_{eff}=-\frac{\lambda _{i1k}^{^{\prime }}\lambda 
_{i2k}^{^{\prime }}}{%
m_{\widetilde{d_k}}^2}\left[ \overline{\left( l_{_iL}\right) 
^c}u_L\mbox{ }%
\overline{s_L}\left( \nu _{iL}\right) ^c\right] .  
\lb{eq:3}
\end{equation}
Thus, the amplitude for the process induced by (\ref{eq:3}) is given 
by

\begin{equation}
M_{R\!\!\!/}=\frac{\lambda _{i1k}^{^{\prime }}\lambda 
_{i2k}^{^{\prime 
}}}{%
m_{\widetilde{d_k}}^2}\stackrel{\_}{l}\left( 1+\gamma _5\right) 
u\mbox{ 
}%
\stackrel{\_}{s}\left( 1-\gamma _5\right) \nu _l.  
\lb{eq:4}
\end{equation}
After Fierz's re-ordering (\ref{eq:4}) takes the form
\begin{equation}
M_{R\!\!\!/}=\frac{\lambda _{i1k}^{^{\prime }}\lambda 
_{i2k}^{^{\prime 
}}}{%
m_{\widetilde{d_k}}^2}\stackrel{\_}{\nu }_l\gamma _\sigma \left( 1-
\gamma
_5\right) l\mbox{ }\stackrel{\_}{s}\gamma ^\sigma \left( 1-\gamma 
_5\right)
u.  
\lb{eq:5}
\end{equation}
Adding (\ref{eq:5}) in SM's contribution, the total amplitude can be 
expressed in the
form

\[
M^{^{\prime }}=\left( \frac{G^{^{\prime }}}{\surd 2}\right) 
\stackrel{\_}{%
\nu }_l\gamma _\sigma \left( 1-\gamma _5\right) 
l\stackrel{\_}{s}\gamma
^\sigma \left( 1-\gamma _5\right) u, 
\]
where $G^{^{\prime }}/\surd 2=\left( G_FV_{us}/\surd 2+\lambda
_{i1k}^{^{\prime }}\lambda _{i2k}^{^{\prime 
}}/8m_{\widetilde{d_k}}^2\right)
.$ In the limit of vanishing lepton masses the branching ratio for a
baryonic decay under consideration is given by

\begin{eqnarray*}
Br\left( B^{}\longrightarrow B^{\prime }\mbox{ }l\mbox{ 
}\overline{\nu 
}%
_l\right) &=&\tau _B\mbox{ }\frac{G^{^{\prime }2}g_V^{2_{}}(0)}{60\pi 
^3}%
\mbox{ }\left( \frac{m_{B^{^{\prime }}}}{m_B}\right) ^{3/2}\Delta 
^5\times \\
&&\left[ 1+3\alpha ^2+\left( \frac{\Delta ^2}{2m_Bm_{B^{^{\prime 
}}}}\right)
\left( 0.96+1.18\right) \alpha ^2\right] ,
\end{eqnarray*}
where $\Delta =m_{B^{^{}}}-m_{B^{^{\prime }}}$, $\alpha =g_A/g_V$ and 
$\tau
_B$ is life time of the parent particle.

Various constraints on the product of $\lambda _{i1k}^{^{\prime 
}}\lambda
_{i2k}^{^{\prime }}$ $\left( i=1,2\right) $ in the units of $\left( 
m_{%
\widetilde{d_k}}/100GeV\right) ^2$ from different baryonic decays are 
listed
in the Tables 1 and 2.

{\bf Table 1} 
\[
\begin{tabular}{|c|c|c|}
\hline
{\bf Process} & {\bf BranchingRatio (Expt)} & $\lambda
_{11k}^{^{\prime }}\lambda _{12k}^{^{\prime }}$ \\ \hline\hline
$\Lambda \longrightarrow Pe^{-}\overline{\nu }_e$ & $<\left( 8.32\pm
0.14\right) \times 10^{-4}$ & $
\begin{array}{c}
<{1.2\ \times 10^{-1}\ (1\ \sigma )\ } \\ 
<{1.3\ \times 10^{-1}\ (2\ \sigma )}
\end{array}
$ \\ \hline
$\Sigma ^{-}\longrightarrow ne^{-}\overline{\nu }_e$ & $<(1.017\pm
0.034)\times 10^{-3}$ & $
\begin{array}{c}
<{8.1\ \times 10^{-2}\ (1\ \sigma )\ } \\ 
<{8.5\ \times 10^{-2}\ (2\ \sigma )}
\end{array}
$ \\ \hline
$\Xi ^{-}\longrightarrow \Lambda e^{-}\overline{\nu }_e$ & $<(5.63\pm
0.31)\times 10^{-4}$ & $
\begin{array}{c}
<{1.1\ \times 10^{-1}\ (1\ \sigma )} \\ 
<{1.2\ \times 10^{-1}\ (2\ \sigma )}
\end{array}
$ \\ \hline
\end{tabular}
\]

\ {\bf Table 2}
\[
\begin{tabular}{|c|c|c|}
\hline
{\bf Process} & {\bf BranchingRatio(Expt)} & $\lambda
_{21k}^{^{\prime }}\lambda _{22k}^{^{\prime }}$ \\ \hline\hline
$\Lambda \longrightarrow P\mu ^{-}\overline{\nu }_\mu $ & $<(1.57\pm
0.35)\times 10^{-4}$ & $
\begin{array}{c}
<{1.6\ \times 10^{-2}\ (1\ \sigma )\ } \\ 
<{5.3\ \times 10^{-3}\ (2\ \sigma )}
\end{array}
$ \\ \hline
$\Sigma ^{-}\longrightarrow n\mu ^{-}\overline{\nu }_\mu $ & $<(4.5\pm
0.4)\times 10^{-4}$ & $
\begin{array}{c}
<{9.6\ \times 10^{-3}\ (1\ \sigma )\ } \\ 
<{1.6\ \times 10^{-2}\ (2\ \sigma )}
\end{array}
$ \\ \hline
$\Xi ^{-}\longrightarrow \Lambda \mu ^{-}\overline{\nu }_\mu $ & $%
<(3.5_{-2.2}^{+3.5})\times 10^{-4}$ & $<{5.0\ \times 10^{-2}\ }$ \\ 
\hline
\end{tabular}
\]
Bounds on new set of product couplings summarized in tables 1 and 2, 
are
within a range obtained for the products $\lambda _{31k}^{^{\prime 
}}\lambda
_{11k}^{^{\prime }}$, $\lambda _{31k}^{^{\prime }}\lambda 
_{21k}^{^{\prime
}} $ \ct{12} from the $\tau $-decays into leptons and mesons and 
$\lambda
_{13k}^{^{\prime }}\lambda _{23k}^{^{\prime }}$, $\lambda 
_{23k}^{^{\prime
}}\lambda _{33k}^{^{\prime }}$, $\lambda _{33k}^{^{_{}\prime }}\lambda
_{13k}^{^{\prime }}$ from the rare decays of $Z$ \ct{13}.

In conclusion, we have considered the effects of the SUSY with 
explicitly
broken $R$-parity via $L$-violation on the hypercharge changing 
semileptonic
decays of certain baryons. Interactions induced by $L$-violating 
operator of
the superpotential put constraints on the products of couplings 
$\lambda
_{11k}^{^{\prime }}\lambda _{12k}^{^{\prime }}$ and $\lambda
_{21k}^{^{\prime }}\lambda _{22k}^{^{\prime }}$.

\end{document}